\documentclass[fleqn,twoside]{article}
\usepackage{espcrc2}
\usepackage{graphicx}
\usepackage[figuresright]{rotating}

\title{Mildly mixed coupled models {\it vs.}~WMAP7 data}

\author{Giuseppe~La~Vacca \& Silvio~A.~Bonometto \address {Physics
Department, Milano--Bicocca University \& I.N.F.N.,
Sez.~Milano--Bicocca\\~~ Piazza della Scienza 3, 20126 Milano, Italy}%
\thanks{Talk given by S.A.B.; his present address is: Physics
Department, Section Astrophysics, Trieste University, Via Tiepolo 11,
34143 Trieste, Italy} }
       
\begin{document}

\begin{abstract}
Mildly mixed coupled models include massive $\nu$'s and CDM--DE
coupling. We present new tests of their likelihood {\it vs.}~recent
data including WMAP7, confirming it to exceed $\Lambda$CDM, although
at $\sim 2$--$\sigma$'s. We then show the impact on the physics of the
dark components of $\nu$--mass detection in $^3$H $\beta$--decay or $0
\nu \beta \beta$--decay experiments.
\vspace{1pc}
\end{abstract}

\maketitle

\section{Spectral distorsions}
Cosmological data (apart $^7$LI abundance) are nicely fitted by
$\Lambda$CDM, a model which however has severe {\it fine tuning} and
{\it coincidence} problems. Here we therefore discuss an alternative
easing these problems: that, symoultaneously, neutrinos ($\nu$) have
mass, and DE is a scalar field $\phi$ self--interacting and
interacting with Cold Dark Matter (CDM). To our knowledge, this is the
only alternative whose likelihood, although marginally, exceeds
$\Lambda$CDM.

\begin{figure}[h!]
\begin{center}
\includegraphics[height=6.cm,width=7.5truecm]{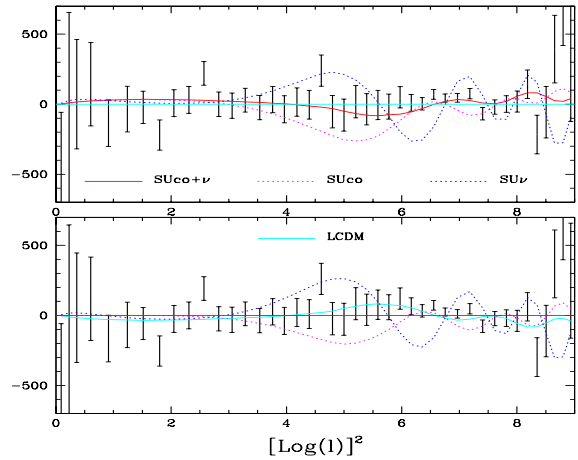}
\end{center}
\vskip-1.truecm
\caption{$\Lambda$CDM and MMC $C_l$ compared. In the upper (lower)
plot $C_l$ are normalized to $\Lambda$CDM (the best fitting SUGRA
model including coupling and massive $\nu$'s). $C_l$ obtained with
either coupling or massive $\nu$'s only are also shown. The error bars
are a sampling of WMAP7 $C_l$ data.}
\label{cmb}
\end{figure}

An energy transfer from CDM to Dark Energy (DE) causes significant
distorsions of $C_l$ and $P(k)$ spectra in respect to $\Lambda$CDM,
but allows DE to be a significant cosmic component since ever;
distortions are also caused by $\nu$ masses, in the range $M_\nu =
\sum_i m_i \sim 1$~eV. These two distorsions tend however to
compensate and compensation allows to fit data better than
$\Lambda$CDM (Figure \ref{cmb} shows this for CMB anisotropy
spectrum).

This yields models including a slight amount of Hot Dark Matter
(typically $\Omega_h \sim 0.01$); they are then {\it Mildly Mixed} and
{\it Coupled} (MMC) models.

\section{Potentials \& coupling}
Among possible CDM--DE couplings \cite{lopez}, we consider the option
arising from Brans--Dickie gravity conformally transformed from the
Jordan to the Einstein frame \cite{amend00}, just allowing for a
generalized self--interaction potential. Then, while
$T^{(c)~\mu}_{~~~~\nu;\mu} + T^{(de)~\mu}_{~~~\, ~~\nu;\mu} = 0\, $,
it is
\begin{equation}
T^{(de)~\mu}_{~~~\, ~~\nu;\mu} = +C T^{(c)} \phi_{,\nu}~,~~
T^{(c)~\mu}_{~~~~\nu;\mu} =- C T^{(c)} \phi_{,\nu}~,
\label{conti1}
\end{equation}
($T^{(c,de)}_{\mu\nu}:$ CDM and DE stress--energy tensors,
$T^{(c,de)}:$ their traces) with a coupling
\begin{equation}
 C = (16 \pi/3)^{1/2} \beta/m_p
\label{beta}
\end{equation}
($m_P:$ Planck mass). Ratra--Peebles (RP, \cite{RP}) or SUGRA
\cite{SUGRA} potentials 
\begin{equation}
V(\phi) = (\Lambda/\phi)^\alpha
~~~~~~~~~~~~~~~~~~~~~~~~ (RP)
\end{equation}
\begin{equation}
V(\phi) = (\Lambda/\phi)^\alpha \exp(4\pi \phi^2/m_p)
~~~ (SUGRA)
\end{equation}
are considered here, for $\phi$ self--interaction, so easing fine
tuning. RP (SUGRA) yields a smooth (fast) dependence of the DE state
parameter on redshift. For both potentials $\Lambda$ will be taken as
a free parameter. Once the density parameter of DE is found, the
valuse of $\alpha$ is also uniquely defined. Both these potentials
also yield a dynamical rise from $\cal O$$(10^{-2})$ to unity of the
DE/CDM ratio, at the eve of the present epoch, so easing coincidence
as well. The natural scale for $C$ is then $\cal O$$(m_p^{-1})$; only
in the presence of $\nu$ masses such range gets consistent with data.

\section{Neutrino mass}
Absolute $\nu$ masses can be measured through double beta decays
($0\nu\beta\beta$) or $^3$H $\beta$ decay \cite{2b,katrin}.

The former process is allowed only if $\nu$'s are Majorana spinors
with mass, yielding
\begin{equation}
m_{\beta\beta}^2 = \sum_i U_{ei}^2 m_i = 
{m_e^2}/{C_{mm} t^{\, 0\nu\beta\beta}_{1/2}}
\end{equation} 
($U_{ei}:$ PMNS $\nu$ mixing matrix; $m_e :$ electron mass;
$t^{0\nu\beta\beta}_{1/2}: $ decay half life). Here, the nuclear
matrix element $C_{mm}$ causes the main uncertainties.

Using $^{76}$Ge, the Heidelberg-Moscow (HM) \cite{baudis} and the IGEX
\cite{igex} experiments gave $t^{0\nu}_{1/2} > 1.9 \times 10^{25} y$
and $t^{0\nu}_{1/2} > 1.6 \times 10^{25} y$, respectively.  However, a
part of the HM theam claims a $t^{0\nu}$ detection yielding
$m_{\beta\beta} \neq 0$ at $> 5 \sigma$'s. At $3\, \sigma$'s, this
KK--claim reads $m_{\beta\beta } = (0.2-0.6)$eV \cite{klapdor}.

The best limits on $m_\beta$ from $^3$H $\beta$--decay come from the
Mainz and Troitsk experiments: $m_\beta <2.0\, $eV, at 95\% C.L.).
The experiment KATRIN \cite{katrin} will soon improve the limit by one
order of magnitude, being able to confirm the KK claim.

This is the range of masses needed to balance DE--CDM coupling, so
yielding MMC models.

\section{Methods \& data}
Here we show results of fits of MMC models to available cosmological
data, performed by using the publicly available code CosmoMC
\cite{cosmomc}. The dataset combinations considered are: (i)
WMAP7+BAO+$H_o$. (ii) WMAP7+BAO+SNIa. (iii) The same data plus the
power spectrum of galaxy surveys.

The following parameters define the model:

\centerline{
\{ $\omega_b, \omega_{c}, \theta, \tau, n_s, \ln
10^{10} A_s, \Lambda, \beta, M_\nu $ \}
}

Here $\omega_{b,c}=\Omega_{b,c} h^2$, $\theta$ is the ratio of the
comoving sound horizon at recombination to its distance, $\Lambda$ is
the energy scale in RP or SUGRA potentials, $\beta$ yields the CDM--DE
coupling, $\tau$ and $A$ have their usual meanings, while $\nu$--mass
differences are neglected.

Results including in datasets SSDS ``data'' \cite{sdss} will be also
shown. Although used also in WMAP7 release \cite{komatsu}, such
``data'' are obtained from observations by exploiting the {\it
Halofit} expressions \cite{halofit} for non--linear spectra. Such
expressions are reliable within the frame of $\Lambda$CDM cosmologies,
but could produce misleading results if the true cosmology is
non--$\Lambda$CDM, as we envisage here. It does not come then as a
surprise that these last results appear much less promising for MMC
cosmologies.

\begin{figure}
\begin{center}
\includegraphics[height=6.cm,width=7truecm]{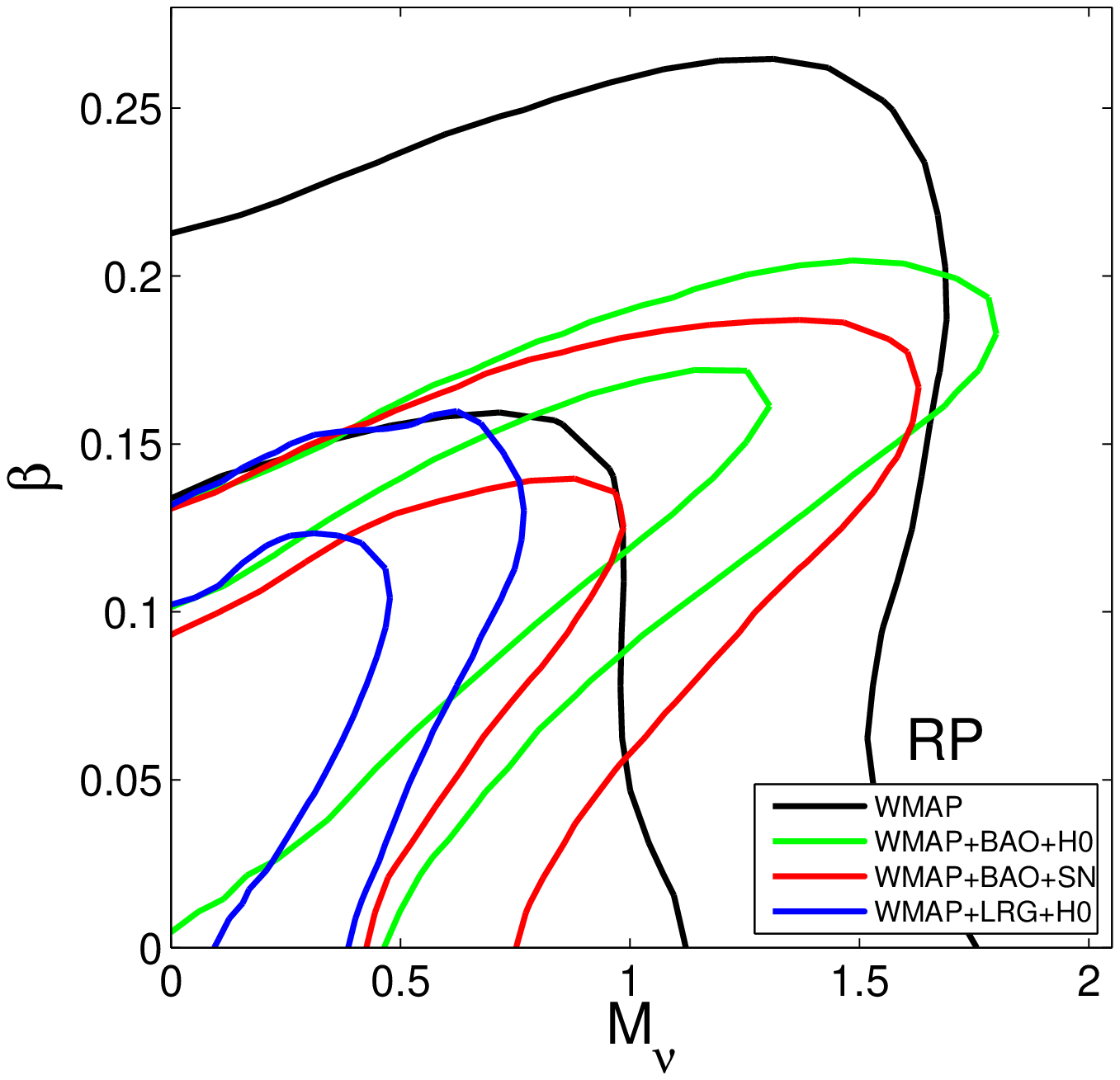}
\includegraphics[height=6.cm,width=7truecm]{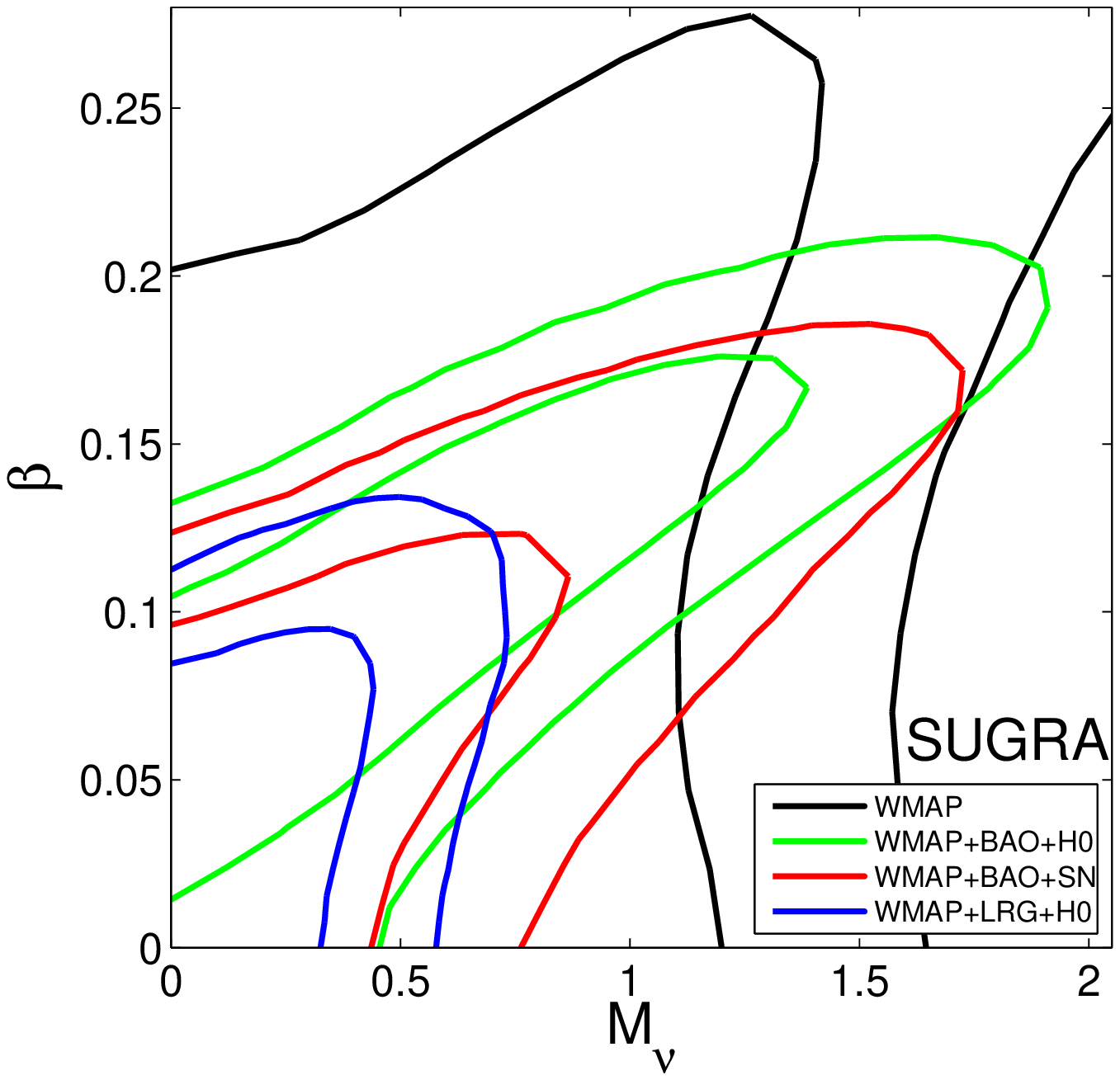}
\end{center}
\vskip-1.truecm
\caption{1-- \& 2--$\sigma$ likelihood contours for various
datasets. The red and green ones, using CMB and BAO data, plus either
$H_0$ or SNIa constraints, confirm the correlation between $\beta $
and $M_\nu$.  WMAP7 data only provide quite loose contours. Blue
contours include SSDS spectral data (LRG). They are reported for the
sake of completeness, but they assume {\it Halofit} spectral
expressions, unsuitable to fit non--$\Lambda$CDM models.  }
\label{elli01}
\end{figure}
\begin{figure}
\begin{center}
\includegraphics[height=6.cm,width=7truecm]{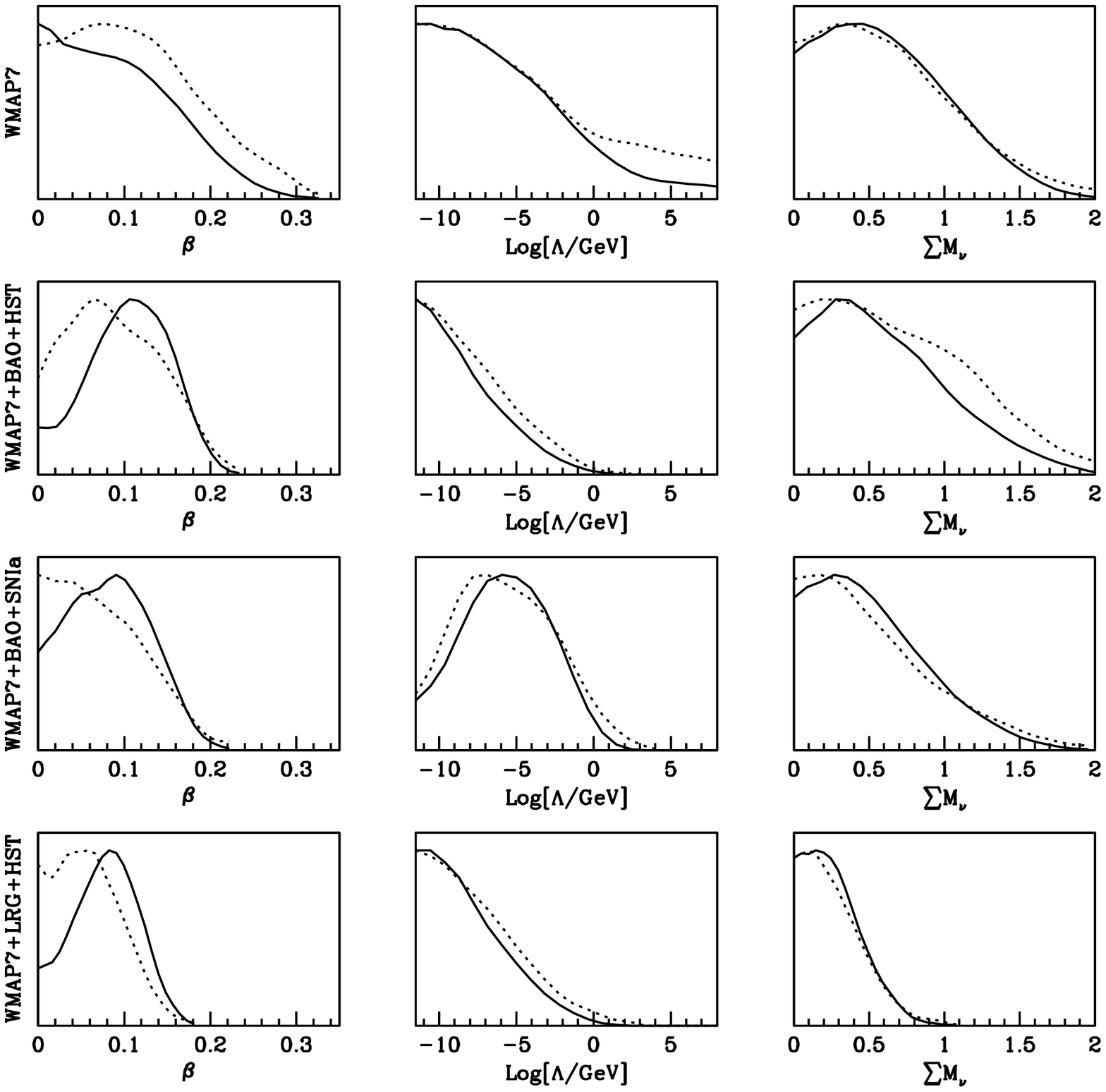}
\vskip .2truecm
\centerline{------------------------------------------------------}
\vskip .6truecm
\includegraphics[height=6.cm,width=7truecm]{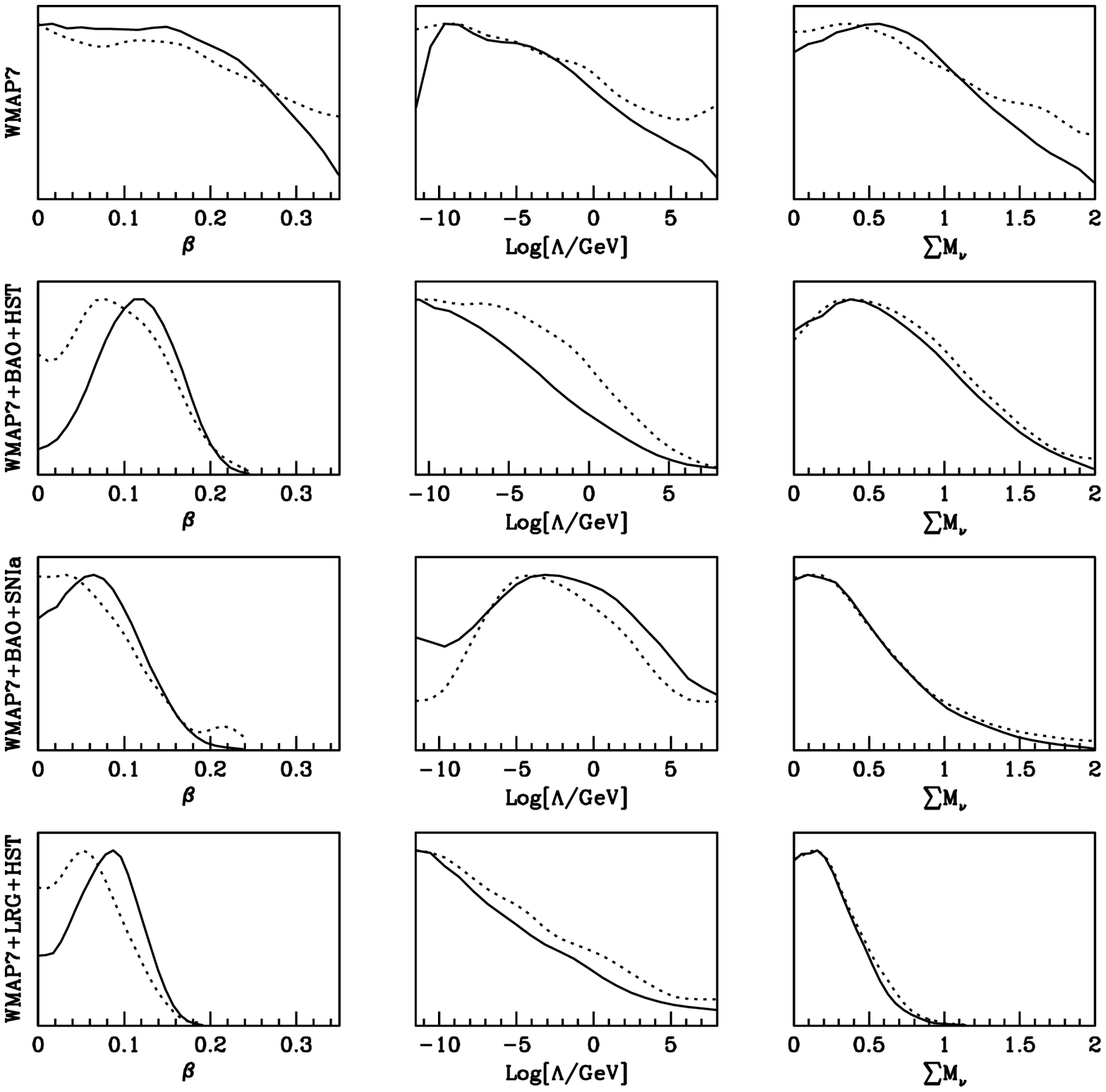}
\end{center}
\vskip-1.truecm
\caption{Likelihood distribution on $\beta$, $\Lambda$, $M_\nu$ for
various datasets (about then see the caption of the previous
Figure). The upper (lower) panel is for RP (SUGRA) potential. Notice
the near--detection of $\beta$ and that $\Lambda$ constraint
are looser than in the absence of coupling.}
\label{1D}
\end{figure}

\section{Results}
In Figure \ref{elli01} we show 1-- and 2--$\sigma$ marginalized
likelihood curves in respect to various datasets (see caption) for RP
and SUGRA potentials. The two panels exhibit just marginal
quantitative shifts and, in the sequel, only RP results will be
reported.

The strong degeneracy between $M_\nu$ and $\beta$ is confirmed,
evident when CMB data are put together with low--$z$ data. If spectral
SDSS data are used, the degeneracy is damped. As previously outlined,
this is not a surprise and calls for an unbiased analysis of the huge
SDSS sample.

In Figure \ref{1D} we show marginalized and average likelihood
distributions on $\beta$ (coupling), $\Lambda$ (energy scale in
potential) and $M_\nu,$ when varying the dataset.

Both $\beta$ and $M_\nu$ plots exhibit a maximum at non--zero values.
The maximum on $\beta$ persists even when spectral data are
considered.

\begin{figure}[h]
\begin{center}
\includegraphics[height=6.cm,width=7truecm]{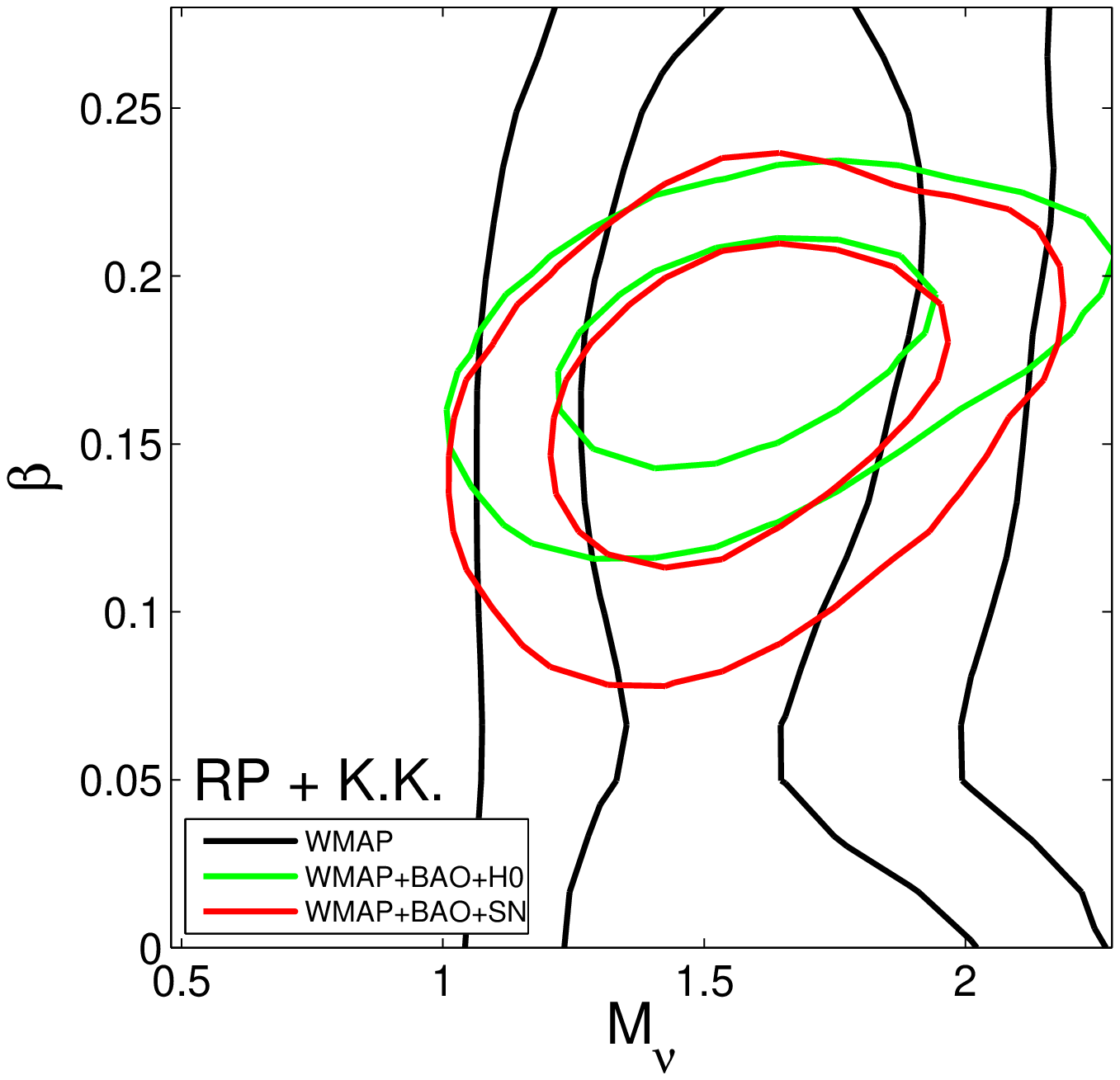}
\includegraphics[height=6.cm,width=7truecm]{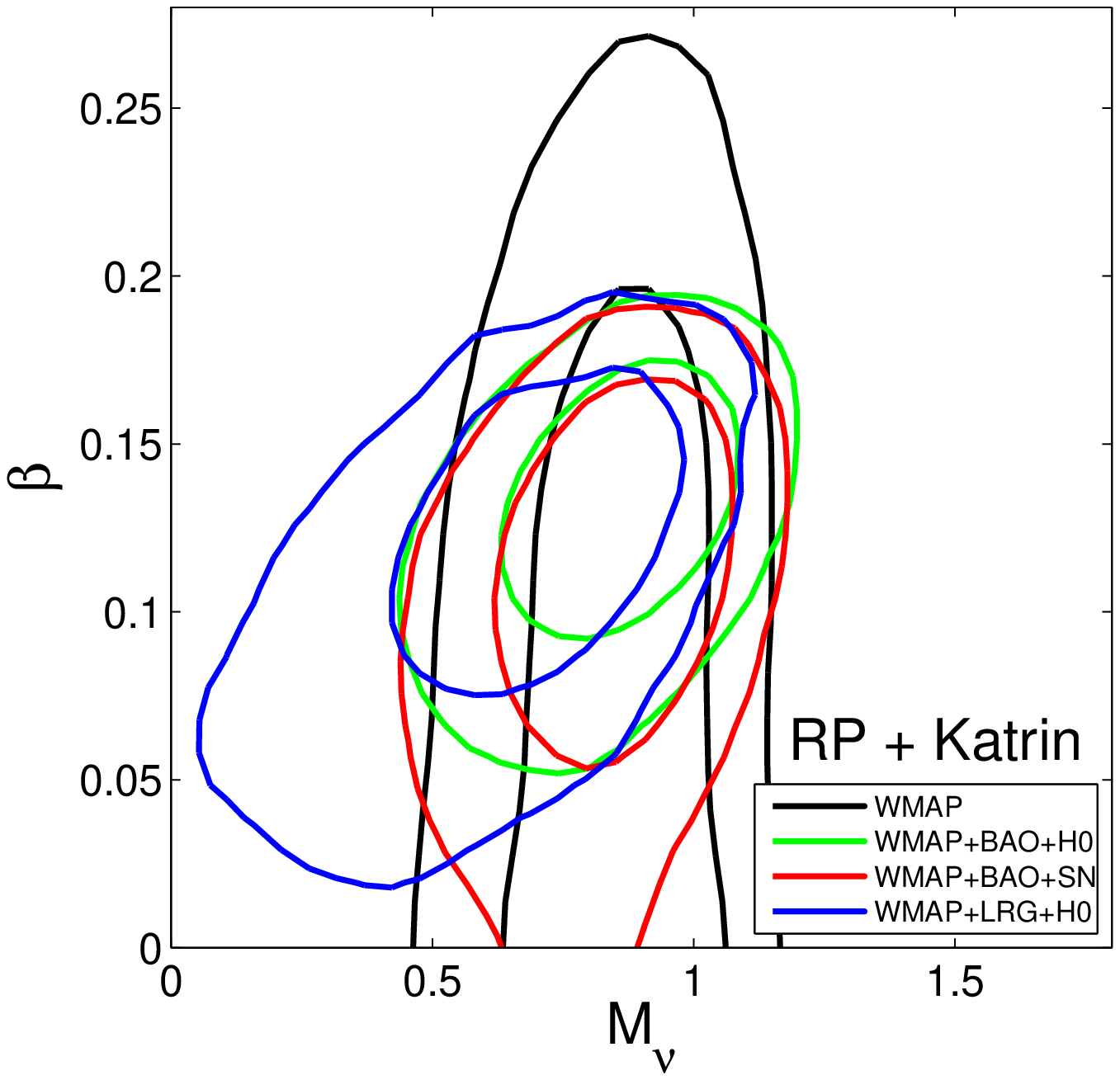}
\end{center}
\vskip-1.truecm
\caption{Likelihood distribution under the assumption of the validity
of the KK--claim ($M_\nu = 1.5$~eV; upper panel) or assuming the
detection of $M_\nu = 0.9$ by KATRIN (lower panel). The plots are
reported for the RP potential only; in the SUGRA case there are no
qualitative differences.}
\label{HM}
\end{figure}

We then studied what effects would arise on cosmological parameters if
the KK-claim is correct (Figure \ref{HM}, upper panel) or the KATRIN
experiment (Figure \ref{HM}, lower panel) leads to $\nu$ mass
detection. The two Figure differ for the range of $\nu$ mass
considered.

In the latter case we assumed $M_\nu \simeq 0.9~$eV and this leads to
the area of top expectation, from cosmological data. On the contrary,
the average KK--claim takes us above such area, although one should
not forget that such claim could be quite consistent with our
``KATRIN'' assumption.

Let us however ooutline that, besides of being consistent with MMC
models, such $\nu$--mass detections apparently imply the discovery
of CDM-DE coupling, possibly at more than 3--$\sigma$'s and a final
overcoming of the $\Lambda$CDM cosmology.

\end{document}